\documentstyle[11pt,aaspp4, flushrt]{article}  

\begin{document}
 
\title{A Hard Tail in the Broad Band Spectrum of
the Dipper XB~1254--690}

\author{R. Iaria\altaffilmark{1},
T. Di Salvo\altaffilmark{1},
L. Burderi\altaffilmark{2},
N. R. Robba\altaffilmark{1}}
\altaffiltext{1}{Dipartimento di Scienze Fisiche ed Astronomiche, 
Universit\`a di Palermo, via Archirafi n.36, 90123 Palermo, Italy}
\altaffiltext{2}{Osservatorio Astronomico di Roma, Via Frascati 33, 
00040 Monteporzio Catone (Roma), Italy}
  
\authoremail{iaria@gifco.fisica.unipa.it}

\begin{abstract}

We report on the results of  spectral analysis
of the dipping source XB~1254--690 observed by the BeppoSAX satellite. 
We find that the X-ray dips are not present during the BeppoSAX observation,
in line with recent RXTE results.
The broad band (0.1--100 keV) energy spectrum is well fitted by a 
three-component model consisting of a multicolor disk blackbody with an 
inner disk temperature of $\sim 0.85$ keV, a comptonized spectrum with an 
electron temperature of $\sim 2$ keV, and a  bremsstrahlung at a temperature 
of $\sim 20$ keV.  
Adopting a distance of 10 kpc and taking into account a spectral hardening
factor of $\sim 1.7$ (because of electron scattering which modifies the 
blackbody spectrum emitted by the disk) we estimated that the inner disk 
radius is
$R_{\rm in} \sqrt{\cos i} \sim 11$ km, where $i$ is 
the inclination angle of the system with respect to the line of sight.  
The comptonized component could originate in a spherical corona or boundary 
layer, surrounding the neutron star, with  optical depth $\sim 19$.  
The bremsstrahlung emission, contributing  $\sim 4\%$
of the total luminosity, 
probably  originates in an extended accretion disk corona with radius 
$\sim 10^{10}$ cm. 
In this scenario we calculated
that the optical depth of this region is $\sim 0.71$
and its mean electron density is $N_e \sim 2.7 \times 10^{14}$ cm$^{-3}$.      
This last component might also be present in other
low mass X-ray binaries, but, because of its low intrinsic luminosity, 
it is not easily observable. 
We also find an absorption edge at $\sim 1.27$ keV with an optical depth of 
$\sim 0.15$. Its energy could 
correspond to the L-edge of Fe XVII, or K-edge of Ne X or neutral Mg.
 
\end{abstract}

 
\section{Introduction}

About 10 Low Mass X-ray Binaries (LMXB) are known to show periodic
dips in their X-ray light curves and most of them also show X-ray burst 
activity. The dip intensities, lengths and shapes  change from source to
source, and, for the same source, from cycle to cycle.
Most of the  characteristics of the dips are consistent with a model in which 
the
central X-ray source is temporarily obscured by material in the 
outer edge of the
accretion disc. This  thicker region could be formed by the impact of the gas 
stream from the Roche-lobe filling companion star
on the outer rim of the accretion disc.
  
Among the dipping sources, XB~1254--690, that also shows type-I X-ray bursts, 
is peculiar because it showed a cessation of its regular dipping activity
(Smale \& Wachter 1999).  This source was observed for the first time by 
HEAO 1 A2 instrument (Mason et al. 1980) when an optical burst of $\sim 20$ s 
duration occurred. 
The optical counterpart to XB~1254--690, GR Mus, was identified by Griffiths 
et al. (1978) to be a faint blue object (V=19.1), implying a 
ratio of X-ray to optical luminosity $L_x/L_{\rm opt} \sim 200$, similar to 
those of other compact X-ray binaries with Population II companions 
(Lewin \& Joss, 1983). 
 
EXOSAT observations showed the presence of type-I X-ray
bursts and led to the discovery of recurrent X-ray dips with a period of 
$3.88 \pm 0.15$ hr (Courvoisier, Parmar \& Peacock 1984; Courvoisier et al. 
1986).  The  duration of the dips was $\sim 0.8$ hrs, representing 
$\sim 20\%$ of the binary cycle, with a reduction of the X-ray flux between
20 and 90\%.  During the dips, X--ray variabilities are present on time 
scales 1--300 s. This complex time structure indicates that the 
obscuring region is clumpy and of variable size.
A modulation with the same period of the X-ray dips was also 
observed in optical light curves, with the optical minimum occurring 
$\sim 0.2$ cycles after the X-ray dip (Motch {\it et al.} 1987).
The presence of dips and the lack of eclipse of the 
X-ray source give a constraint on the inclination angle of the source that 
is between 65$^{\circ}$ and 73$^{\circ}$ (Courvoisier et al. 
1986, Motch {\it et al.} 1987).
X-ray bursts and X-ray dips were also present in a Ginga observation 
performed on August 1990 (Uno et al. 1997).
On the other hand the dips were not present in a RXTE observation in 1997 
between April 28 and May 1 (Smale \& Wachter, 1999). 
The cessation of the dipping activity
indicates that the angular size of the disk edge (at the impact point 
with the accretion stream)  decreased from $17^\circ-25^\circ$ to
less than $10^\circ$ (Smale \& Wachter, 1999). There is no evidence of a 
simultaneous variation of 
the accretion rate or the location of the outer rim of the accretion disk
(Smale \& Wachter, 1999). 

The energy spectrum of XB~1254--690 in the 1--10 keV energy range, 
observed by {\it EXOSAT}, was fitted by thermal bremsstrahlung with 
$kT = 5.5$ keV, and the corresponding flux was $\sim 6 \times
10^{-10}$ ergs cm$^{-2}$ s$^{-1}$ (Courvoisier et al. 1986), 
although other models could not be excluded.  
The spectrum obtained by {\it Ginga} 
in the 2--35 keV range was fitted by a blackbody plus
a power law with exponential cutoff, or by a multicolor disk blackbody 
plus a comptonized blackbody, where the electron temperature of 
the hot plasma is fixed at 100 keV and the optical depth is $\sim 0.2$ (Uno et
al. 1997).  Neither model gave acceptable 
residuals. The fits were partially improved including extra components like 
a Gaussian line centered at 6.3--6.7 keV. 

We report here the results of a broad band (0.1--100 keV) spectral analysis
performed on data from the BeppoSAX satellite.  The broad 
energy band of BeppoSAX narrow field instruments allows us to well constrain 
the spectrum of this source, and to detect the presence of a high energy 
component. 

\section{Observations and Light Curve}

The Narrow Field Instruments (NFI) on board BeppoSAX satellite
(Boella et al. 1997) observed XB~1254--690 on 1998 December 
22 and 23, for a total exposure time of 45 ks.  
The NFIs are four co-aligned instruments which cover more
than three decades of energy, from 0.1 keV up to 200 keV, with good
spectral resolution in the whole range. The LECS (operating in the range 
0.1--10 keV) and the MECS (1.8--10 keV) have
imaging capabilities with field of view (FOV) of $20'$ and $30'$ radius, 
respectively. We selected data for scientific analysis 
in circular regions, centered on the source, of $8'$ and $4'$ radii for  
LECS and MECS, respectively. The background subtraction
was obtained using blank sky observations in which we extracted the 
background spectra in regions of the FOV similar to those used for 
the source. HPGSPC (7--60 keV) and PDS (13--200 keV) are non-imaging 
instruments, because their FOVs, of $\sim 1^\circ$ FWHM, are 
delimited by collimators. The background subtraction for these instruments is, 
generally, obtained using the off-source data accumulated during the rocking 
of the collimators.  However, during this observation, the HPGSPC collimator
did not execute the rocking, neither data during the Earth occultation
were available, because of the high latitude of the source, 
 to do a proper subtraction of the background.
Therefore we could not use the HPGSPC data. 
The energy range used in the spectral analysis for the other NFIs 
is: 0.12--4 keV for the LECS, 1.8--10 keV
for the MECS and 15--100 keV for the PDS.
Different normalizations of the NFIs are considered in the spectral fitting
by including in the model normalizing factors, fixed to 1 for the MECS, and 
kept free for the other instruments.  
We rebinned the energy spectra in order to have at least 25 counts/channel 
and to sample the instrument resolution with the same number of channels 
at all energies\footnote{see the BeppoSAX cookbook at
http://www.sdc.asi.it/software/index.html}. 
 
In Figure 1 we plotted the XB~1254--690 light curve in the energy bands
1.8--10 keV (MECS data, upper panel), and 15--100 keV (PDS data, lower panel)
using a bin time of 256 s. 
During our observation the light curves do not show significant variability.  
In particular, the count rate ranges between 8 and 9 counts/s 
in the MECS, without displaying the regular reduction of flux 
(generally $\sim 20-90\%$) that is typical of this source during the dipping
activity. The total observation time is $\sim 90$ ks, and therefore the 
BeppoSAX observation should contain 6 complete orbital cycles. Considering that
the dips are usually observed at every orbital cycle (only one dip has been 
skipped before the RXTE observation, see Smale \& Wachter 1999), we should 
observe 6
dips in the BeppoSAX light curve. None of them is visible in Figure 1.
We also divided the MECS data into two energy bands, soft (1.8--4 keV) and 
hard (4--10 keV), and calculated the corresponding hardness ratio.
The hardness ratio is constant implying that no spectral variations are
present. This confirms the absence of dips during
our observation, because dips are usually associated with an increase of the
low energy absorption, and therefore with an anti-correlation between
hardness ratio and intensity.
This cessation of the dipping activity was also observed in a previous RXTE 
observation (1997 May, Smale \& Wachter 1999), 
indicating that the cessation of the dipping activity
might be a long-lived phenomenon. 

The observed average flux of the source in the 1--10 keV energy range
is $9.1 \times 10^{-10}$ ergs cm$^{-2}$ s$^{-1}$, which, adopting a distance 
of 10 kpc (Courvoisier {\it et al.} 1986, Motch {\it et al.} 1987), 
corresponds to an unabsorbed luminosity of $1.1 \times 10^{37}$ ergs/s.
This is compatible with the previously reported 
persistent luminosity of the X-ray source of $1 \times 10^{37}$ ergs/s 
in the same energy band 
 (Courvoisier {\it et al.} 1986, Motch {\it et al.} 1987). 
The observed flux of the source in the 0.1--100 keV energy range
is $1.4 \times 10^{-9}$ ergs cm$^{-2}$ s$^{-1}$, which
corresponds to an unabsorbed luminosity of $1.7 \times 10^{37}$ 
ergs/s.

Note that  the diffuse X--ray emission from the thin disk 
surrounding the Galactic midplane (the so-called Galactic ridge), could
affect the background of sources near the Galactic plane. 
Valinia \& Marshall (1998) reported RXTE measurements of the diffuse
X-ray Galactic emission in different Galactic regions. For latitudes
$1.5^{\circ}<|b|<4^{\circ}$ and longitudes $|l|<15^{\circ}$, they find that
the diffuse Galactic average surface brightness is 
$8.1 \times 10^{-8}$ ergs cm$^{-2}$ s$^{-1}$ sr$^{-1}$ in the 2--10 keV band, 
which corresponds to a flux of
$1.1 \times 10^{-13}$ ergs cm$^{-2}$ s$^{-1}$ in the MECS FOV. 
In the same range the flux of XB 1254-690 is 
$6.9 \times 10^{-10}$ ergs cm$^{-2}$ s$^{-1}$, several orders of 
magnitude above the diffuse Galactic emission. At higher energies 
(10--60~keV) the
Galactic ridge emission shows a power-law tail with photon index 
$\sim 1.8$, the flux of which is $3.3 \times 10^{-11}$ ergs cm$^{-2}$ s$^{-1}$
in the PDS FOV ($\sim 1.3^\circ$~FWHM). This is comparable to the flux, in the 
same range, of the bremsstrahlung hard-tail alone (see below), which results 
$2.0 \times 10^{-11}$ ergs cm$^{-2}$ s$^{-1}$.  However,
considering that the diffuse Galactic emission drops rapidly with 
increasing latitude, 
the relative flux is lower in the region of XB 1254--690, which corresponds 
to $b\simeq -6.4^{\circ}$ and $l\simeq -56.5^{\circ}$. 
In fact, for $|b| > 1^\circ$ the Galactic ridge flux has a broad distribution,
that is a Gaussian function with $FWHM = 4-5^\circ$ (see Fig. 2b in
Valinia \& Marshal 1998). We calculated that the flux in the 10--60~keV band
as a function of the latitude $b$ is:
\begin{equation}
\frac{d F(b)}{d b} = 4.41 \times 10^{-8} exp\left[-\frac{b^2}{5.79}\right]
\end{equation}
in ergs cm$^{-2}$ s$^{-1}$ sr$^{-1}$ deg$^{-1}$, where we assumed
$FWHM = 4^\circ$. For $b=6.4^\circ$ the expected Galactic ridge flux
is $\sim 1 \times 10^{-13}$ ergs cm$^{-2}$ s$^{-1}$ in the PDS FOV, much lower
than the flux in the bremsstrahlung hard tail.

\section{Spectral Analysis}

A simple model consisting of a single component, like a power law multiplied 
by an exponential cutoff, or a comptonized spectrum 
(hereafter {\it Compst}, Sunyaev \& Titarchuk 1980), plus  
photoelectric absorption by cold matter,
was not sufficient to fit the spectrum in the whole energy band. In fact  a
soft excess and  a tail 
at energies higher than 20 keV are present in the residuals.
Using the Compst model (see Table 1, Model 1) we obtained 
$\chi^2/d.o.f. = 304/180$.  
Using the more recent {\it Comptt} model (Titarchuk et. {\it al}, 1994),
instead of {\it Compst}, we obtain a seed-photon temperature $kT_0 \sim 0.1$
keV, close to the limit of the low energy instrument. Because of low 
energy absorption by the interstellar medium and
the low statistics at  these energies, the seed-photon temperature has
a large error and can not be determined reliably. 
Therefore we decided to use the  {\it Compst} model, which, 
in this case (low temperatures and high optical depths), gives a 
fit equivalent to that with Comptt, with similar values of the electron 
temperature and the optical depth. 
The same choice was made by Barret et {\it al}. (2000) for the source 
KS 1731--360.

We added a multicolor disk blackbody (hereafter {\it MCD}, Mitsuda et  
{\it al.} 1984) to the previous model to fit the soft excess 
(see Table 1, Model 2).  An F-test indicates that the 
addition of this component improved the fit with high statistical 
significance (probability of chance improvement $1.3 \times 10^{-8}$). 
The addition of MCD did not eliminate the feature present 
at the high energy. The broad band (0.1--100 keV) spectrum is shown in 
Figure 2 (upper panel), and the residual (in units of $\sigma$) with respect 
to this model are shown in Figure 2 (middle panel), where the high energy 
feature above 20 keV is clearly visible. 

This hard excess can not be fitted by the two-component models that 
we tried.
Therefore we added an other component to the previous model. 
The addition of a bremsstrahlung component fits the data at energies
higher than 20 keV, reducing the  $\chi^2/d.o.f.$ from  
$248/178$  to $221/176$ (Table 1, Model 3).  
The improvement of the fit is significant
at $\sim 99.996 \%$ confidence level.  However other models for this
high energy component, such as a power law with high energy cutoff, cannot
be excluded.  We note that the hard excess can not be fitted by 
a reflection model: using blackbody plus {\it Pexriv} model 
(Magdziarz \& Zdziarski, 1995) we obtain $\chi^2/d.o.f.= 245/177$. 
Finally the addition of an absorption edge at 1.26 keV 
gave a further improvement of the fit at $\sim 99.98 \%$ confidence 
level. The residuals (in unit of $\sigma$) corresponding to the 
best fit model are shown in Figure 2 (lower panel), and the corresponding 
values of the parameters are shown in Table 1 (Model 4). 
In Figure 3 we show the unfolded spectrum 
and the best fit model.

We obtain a hydrogen equivalent column $N_{\rm H}
\simeq 3.5 \times 10^{21}$ cm$^{-2}$, in agreement with previous results
(EXOSAT observation,  Courvoisier {\it et al.} 1986).  
The soft component is consistent
with emission from an accretion disk with an inner disc temperature 
$\sim 0.85$ keV, and an inner radius $R_{\rm in} \sqrt{\cos i} \simeq 3.9$ km
for a distance of 10 kpc. 
The comptonized component originates from the Comptonization of soft seed 
photons in a hot ($kT_{\rm e} \sim 2$ keV) corona of moderate optical depth 
($\tau \sim 19$ for a spherical geometry). The high energy spectrum is 
fitted by a thermal bremsstrahlung corresponding to a 
temperature of $\sim 20$ keV and a luminosity of $\sim 6.2 \times 10^{35}$
ergs/s, that is $\sim 4\%$ of the total observed luminosity. 
Finally, we find an absorption edge at an energy of $\sim 1.27$ keV, with
an optical depth of 0.15. 
On the other hand we do not observe an iron K-edge at 7.1 keV, with an
upper limit, at $90 \%$ confidence level, on the optical depth of $\sim 0.033$.
Also, we do not observe an iron emission line at 6.4--6.7 keV.  
The upper limit on the equivalent width is 18 eV for 
a narrow line at 6.4 keV ($FWHM = 0$ keV) and 15 eV for an iron line at 
6.7 keV with $FWHM = 1$ keV.

\section{Discussion}

We analyzed data of the dipping source XB~1254--690 from a BeppoSAX 90 ks
observation in the energy range 0.1--100 keV.  The light curve does not 
show periodic modulations or dipping
episodes. On the contrary, it is quite constant with a count rate around
10 counts/s in the MECS. This source has  shown periodical dips
every orbital period of $\sim 3.9$ hrs  (Courvoisier et al. 
1986).  During these dips, which last about 0.8 hr per cycle, 
usually the 1--10 
keV flux is reduced up to 95\%.  The cessation of the dipping activity was 
observed for the first time in a RXTE observation in 1997, from April 28 to 
May 1 (Smale \& Wachter 1999), and it was ascribed to a reduction of the 
scale height of
the disk to less than $10^\circ$. Our BeppoSAX data indicate that 
cessation of the dipping activity was also present in December 1998. Therefore
it does not seem to be a short-lived phenomenon. Moreover it is not associated
with evident changes in the X-ray flux, because the persistent luminosity
during the BeppoSAX observation, $\sim 10^{37}$ ergs/s, is not different from
previous measurements (Courvoisier {\it et al.} 1986, Motch {\it et al.} 1987).

We performed a spectral  analysis  of XB~1254--690, and the results are 
discussed in the following.  
For the equivalent absorption column  $N_H$ we obtained a value of  
$0.35 \times 10^{22}$ cm$^{-2}$ in agreement with previous observations 
(Courvoisier {\it et al.} 1986). 
For a distance to the source of 10 kpc (Courvoisier {\it et al.} 1986, 
Motch {\it et al.} 1987) the visual extinction in the 
direction of XB~1254--690 is $A_v=3.1 \pm 1.1$ mag (Hakkila et al. 1997). 
Using the observed correlation between visual extinction and 
absorption column (Predehl \& Schmitt 1995) we find 
$N_H=\left(0.55 \pm 0.20 \right) \times 10^{22}$ cm$^{-2}$,
similar to the value obtained by the spectral fit.
   
A three-temperature model plus an absorption edge are  needed to fit 
the broad band spectrum.  The first
component, in order of increasing temperature, 
is a soft emission that can be described by  a multicolor 
disk blackbody.
We obtain an inner disk temperature of $\sim 0.85$ keV and an inner radius 
of the disk of $R_{\rm in} \sqrt{\cos i} \simeq 3.9$ km. 
This source is a dipper so the inclination angle of the normal 
to the disk plane with respect to the line of sight should be between 
$65^\circ$ and $73^\circ$ (Courvoisier et al. 1986, Motch et al. 1987). 
Assuming an inclination angle $i$ of $65^\circ$ we obtain 
$R_{\rm in} \sim 6$ km.
Moreover, because the electron scattering can modify the spectrum
(Shakura \& Sunyaev 1973; White, Stella \& Parmar 1988), we have 
corrected the value of the inner radius and its temperature for this effect.
The measured color temperature is related to the effective temperature of the 
inner disk by the relation $T_{col}=f T_{eff}$, where $f$ is the spectral 
hardening factor. It 
was estimated to be $\sim 1.7$ for a luminosity $\sim 10 \%$ of the Eddington
limit (Shimura \& Takahara 1995; Merloni et al., 2000). Applying this
correction to the values of
$T_{in}$ and $R_{\rm in} \sqrt{\cos i}$ reported in Table 1 (Model 4)
we obtain an 
effective temperature of $\sim 0.5$ keV which corresponds to an inner radius
$R_{eff}\sqrt{\cos i} \simeq 11$ km. For $i=65^\circ$ we obtain  
$R_{eff} \simeq 17$ km. 
We identify this soft component with emission from the accretion disk. In fact
given the high inclination of the system, we expect that the emission
of the neutron star will not be directly observed because it is probably 
reprocessed by the corona that originates the comptonized component.

From the soft component, interpreted as emission from the accretion disk,
we can calculate the mass accretion rate of the system. The luminosity of this
component is $L_{\rm disk}=2 \pi D^2 F/\cos i$, where D is the distance to
the system and F is the flux measured from the Earth. Using the values
reported in Table 1 (Model 4), we obtain $L_{\rm disk} \sim  
10^{36}/\cos i$ ergs/s. This is also equal to the potential energy of the 
matter released at the 
inner radius of the disk, $L_{\rm disk}=G M \dot{M}/(2R_{\rm in})$. The 
accretion rate we derive from this relation is 
$\dot{M} \simeq 1.2 \times 10^{16}(\cos i)^{-3/2}$ g/s 
$\sim 4.4 \times 10^{16}$ g/s. 
The corresponding total luminosity is $\sim 8 \times 10^{36}$ ergs/s, 
similar to the observed luminosity of $\sim 1.7 \times 10^{37}$ ergs/s. 

The second component of our model is a comptonized spectrum, produced in
a hot ($k T_{\rm e} \sim 2$ keV) region of moderate optical depth 
($\tau \sim 19$, for a spherical geometry) probably surrounding the 
neutron star. 
Adopting the geometry described above one would expect a Compton reflection
of the hard spectrum emitted by the corona from the accretion disk.  This 
would produce a bump in the energy spectrum as well as an emission line at 
$\sim 6.4$ keV and an edge at $\sim 7$ keV from neutral iron ({\it e.g.} 
George \& Fabian 1991; Matt, Perola, \& Piro 1991). 
This reflection signatures are not present in XB~1254--690. Indeed the upper 
limit to the equivalent width of an iron line at $\sim 6.4$ keV is $\sim 15$
eV. 
A possible reason could be the high inclination, that reduces the amount
of reflection seen. Moreover, because of the low temperature of the corona, 
most of the hard photons have energies below 10 keV, and therefore they will 
be photoelectric absorbed by the matter in the disk and re-emitted at the 
temperature of the disk rather than Compton scattered.

The comptonizing region could be identified with an inner, optically thick
($\tau \sim 19$) corona probably surrounding the neutron star. Its relatively 
low electron temperature ($\sim 2$ keV) is determined by Compton cooling
due to the soft emission of the neutron star intercepted by the corona.
Another possible origin of this comptonized spectrum could be the boundary
layer between the neutron star and the accretion disk. In neutron star 
binaries an expanded, hot, low density boundary layer should exist, which
cools by inverse Compton scattering of the soft photons emitted by the
neutron star surface (Popham \& Sunyaev 2000). Popham \& Sunyaev (2000)
predict  
that for an accretion rate around $\sim 10^{17}$ g/s (corresponding to the
the observed luminosity of XB~1254--690), the electron 
temperature  of the boundary layer
is around 1.5-2 keV, in agreement with the value obtained from our analysis.
Moreover, the model predicts that the boundary layer has a radius of
$\sim 13$ km, similar to the inner radius of the accretion disk that
we obtain from the data ($\sim 17$ km). In this scenario the scale height 
of the boundary layer is $\sim 6-10$ km, and therefore it can easily occult
the emission of the neutron star from direct observation given the high
inclination ($\sim 65^\circ$) of the system. This is in agreement with
our identification of the observed soft component with emission from the 
accretion disk, instead of emission from the neutron star surface.

The spectrum described above is similar to that of other type-I X-ray bursters
(see {\it e.g.} Guainazzi et al. 1998, Barret et al. 2000, 
Di Salvo et al. 2000).  For these sources the electron
temperature of the scattering region is between 3 and 28 keV and the
optical depth between 3 and 10, with higher temperatures corresponding
to lower optical depths. In XB~1254--690 the temperature of 
the scattering region is $\sim 2$ keV and the optical depth is $\sim 19$. 
An interesting feature, in XB~1254--690, is the presence of
a hard component, that can be fitted with 
a high temperature ($\sim 20$ keV) bremsstrahlung. This is a low
luminosity ($L \sim 6.2 \times 10^{35}$ ergs/s) component, contributing 
only $\sim 4\%$ of the total luminosity.  This can explain why we observe
such a component in a high inclination source. In fact in this case the 
luminosity of other components, {\it e.g.} the disk, is lowered by the 
geometrical factor $\cos i$. Also the low temperature of the comptonized 
spectrum enhances the possibility to observe such a component.

Assuming  the bremsstrahlung normalization as:
\begin{equation}
N_{\rm bremss} = \frac {3.02 \times 10^{-15}}{4 \pi D^{2}} 
N^{2}_{e} V
\end{equation}
where D is the distance to the source in cm, N$_e$ is  the 
electron density (cm$^{-3}$) and
V is the volume of the  bremsstrahlung emitting region, and
using the value of $N_{\rm bremss}$ reported in Table 1 (model 4), we
obtain an emission measure of $N^{2}_{e} V \simeq 1.8 \times 10^{58}$
cm$^{-3}$. This value requires large emission volumes for reasonable electron
densities.
Considering an optical depth $\tau \sim 1$ and a spherical
volume for the bremsstrahlung emitting region, the condition above gives 
a radius of $\sim 0.2 \times 10^5$ km, that
is an order of magnitude smaller than the accretion disk radius 
(that is $3.6 \times 10^5$ km as reported by Courvoisier et al. 1986).
 
In neutron star binaries an accretion disk corona (hereafter ADC) should be
present ({\it e.g.} White \& Holt 1982), probably formed by evaporation of 
the outer layers of the disk illuminated by the emission of the central 
object. The partial eclipses observed in the light curves of sources
viewed almost edge-on indicates that these ADCs are extended, with radii
that can be a fraction of the accretion disk radius. We suggest that
the high temperature bremsstrahlung component is emitted in an extended,
optically thin ADC. In this hypothesis
the volume of the bremsstrahlung emitting region can be written as 
$V = 4 \pi R_c^3/3$; the ADC radius $R_c$, according to White \& Holt (1982), 
can be obtained by $R_c \simeq (M_{\rm NS}/M_{\odot})T_7^{-1}R_{\odot}$, 
where $M_{\rm NS}$ is the mass of the compact object, 
$M_{\odot}$ and $R_{\odot}$ are mass and radius of the Sun, 
and $T_7$ is the ADC temperature in units of $10^7$ K. Substituting $\sim 20$ 
keV for the ADC temperature and 1.4 $M_{\odot}$ for the neutron 
star mass we obtain  $R_c \simeq 0.4 \times 10^5$ km, similar to 
the value obtained above considering an optically thin emitting region.
Assuming $N_{\rm bremss}$ as reported in Table 1 (Model 4), 
we obtain $N_e \sim 2.7 \times 10^{14}$ $cm^{-3}$.  
We also calculated the optical depth of this region. Using  
the Thomson cross section, the ADC radius as geometrical depth and the 
density of the region as reported above,  we obtain $\tau = 0.71$.

Finally, in the spectrum of XB~1254--690 an absorption edge, at the energy of 
$\sim 1.26$ keV with an  optical depth of $\sim 0.15$, 
is also present. Its energy is compatible with the energy of K-edges
of Ne X or neutral Mg, or with the L-edge of Fe XVII. Notably no other edges
are present in the spectrum. In particular we do not observe an iron K-edge
at $\sim 7$ keV, with an upper limit on the optical depth of 0.033. A possible
reason could be the softness of the comptonized spectrum. Its temperature 
is $\sim 2$ keV, and hence the spectrum fastly decreases at higher energies,
reducing the number of photons at $\sim 7$ keV that would be able to produce
the iron K-edge. The same results are also obtained for another high 
inclination LMXB, X 1822--371 (Parmar {\it et al.} 2000). In that case an
edge at $\sim 1.33$ keV, compatible with the energy we measure in XB~1254--690,
is required in both the BeppoSAX and ASCA spectra.  

\section{Conclusions}

We analyzed data from a BeppoSAX observation of XB~1254--690 performed in
1998 December  22 and 23. Neither X--ray bursts nor X--ray dips 
were  present during this observation. The energy spectrum is described
by a multicolor disk blackbody, with an inner temperature of $\sim 0.85$ keV 
and an inner disk radius of $R_{\rm eff} \sqrt{\cos i}\sim 11$  km, a 
comptonized spectrum, and bremsstrahlung emission. 
The comptonized spectrum is produced in a relatively cool ($\sim 2$ keV) 
region of moderate optical depth ($\tau \sim 19$ for a spherical geometry), 
probably surrounding the neutron star. This region could be identified with
an inner optically thick corona or with the boundary layer between the 
accretion disk and the neutron star.  The presence of reflection signatures
is not detected with high statistical significance probably because of the 
high inclination of the system. 
The bremsstrahlung component, probably produced in an extended, optically 
thin ADC, has a temperature of $\sim 20$ keV. The corresponding emission 
region has a mean electron density $N_{\rm e} \sim 2.7 \times 10^{14}$ 
cm$^{-3}$, and an optical depth $\tau \sim 0.71$. 
We observe an absorption edge at $\sim 1.27$ keV, but the nature of this edge 
is not clear, given that we do not observe other edges in the spectrum.
We do not observe an iron emission line at 6.4--6.7 keV, with an upper
limit on the equivalent width of 18 eV for the iron line at 6.4 keV and
15 eV for the iron line at 6.7 keV.  

\acknowledgments
We want to thank Dr. A.~Segreto for his useful suggestions and technical
support, and Dr. V.~Teresi for useful discussions.
This work was supported by the Italian Space Agency (ASI), by the Ministero
della Ricerca Scientifica e Tecnologica (MURST).

\clearpage

\section*{TABLE}

\begin{table}[th]
\begin{center}
\footnotesize 
\caption{Results of the fit of the XB~1254--690 spectrum in the energy band 
0.12--100 keV, with a multicolor disk blackbody (MCD), a 
comptonized spectrum modeled by Compst and a bremsstrahlung component
(Bremss).
Uncertainties are at the 90\% confidence level for a single parameter.
In the MCD model the parameter $R_{\rm in} (\cos i)^{1/2}$ is 
calculated assuming a distance to the source of 10 kpc.  
$k T_{\rm e}$ is the 
electron temperature, $\tau$ is the optical depth of the 
scattering cloud. The Compst normalization, N$_{\rm comp}$, is defined
as in XSPEC v.10.  $k T_{\rm bremss}$ is the bremsstrahlung temperature 
and N$_{\rm bremss}$ is defined as in the text.}
\label{tab1}
\begin{tabular}{l|c|c|c|c} 
\tableline \tableline
                  & Compst   & Compst + MCD & Compst + MCD & Compst + MCD   \\
                  &          &     &   Bremss   & Bremss + Edge           \\
 Parameter        & (Model 1)&(Model 2)&(Model 3)&(Model 4) \\
\tableline
$N_{\rm H}$ $\rm (\times 10^{22}\;cm^{-2})$
& $0.404 \pm 0.011$ & $0.362 \pm 0.030$ &
$0.291^{+0.033}_{-0.046}$& $0.348 \pm 0.049$   \\

$k T_{\rm in}$ (keV)
& --& $1.92^{+0.14}_{-0.45}$ &
$0.98^{+0.14}_{-0.07}$&$0.855^{+0.213}_{-0.091}$   \\

$R_{\rm in}$ $\sqrt{\cos i}$ (km)
& -- & $1.31^{+0.43}_{-0.09}$ &
$3.7^{+1.6}_{-1.0}$& $3.9^{+2.0}_{-1.6}$    \\

$k T_{\rm e}$ (keV)
& $1.953 \pm 0.027$ & $2.83^{+0.78}_{-0.63}$ &
$1.86 \pm 0.10$ & $ 1.864^{+0.072}_{-0.104}$   \\

$\tau$
& $16.42 \pm 0.23 $ & $11.8^{+4.1}_{-2.5}$ &
 $20.9^{+8.2}_{-2.5}$ & $18.7^{+4.1}_{-2.2}$   \\

N$_{\rm comp}$ $(\times 10^{-2})$
& $20.41 \pm 0.34$ & $13.5 \pm 1.0$ &
$10.7^{+2.7}_{-5.3}$& $13.8^{+2.6}_{-5.4}$    \\

$k T_{\rm bremss}$ (keV)
& -- & -- &
$21^{+26}_{-11}$ & $20.2^{+26.7}_{-10.8}$   \\

N$_{\rm bremss}$ $(\times 10^{-3})$
& -- & -- &
$5.0^{+16}_{-3.0}$&$5.1^{+16.9}_{-3.1}$ \\

E$_{\rm edge}$ (keV)
& -- & -- & -- & $1.267^{+0.062}_{-0.053}$  \\

$\tau_{\rm edge}$
& -- & -- & -- & $0.154 \pm 0.059$  \\ 

$\chi^2$/d.o.f.
& 304/180 & 248/178 &
221/176 & 200/174\\
\tableline
\end{tabular}
\end{center}
\end{table}

\clearpage
 
\section*{FIGURE CAPTIONS}
\bigskip

\noindent
{\bf Figure 1}:  Upper panel: XB~1254--690 light curve in the energy band 
1.8--10 keV (MECS data). Lower  panel: XB~1254--690 light curve in the energy 
band 15--100 keV (PDS data). The bin time is 256 s.
\\
{\bf Figure 2}: Upper panel: Count spectrum of XB~1254--690 (0.1--100 keV)
and the model, consisting of {\it MCD} and {\it Compst}, reported  in Table 
1 (Model 2).   Middle panel:  Residuals in units of $\sigma$ 
with respect this model.  Lower panel:  Residuals in units of $\sigma$ with 
respect to the best fit model, consisting of {\it MCD}, {\it Compst}, 
bremsstrahlung and absorption edge, reported in Table 1 (Model 4).
\\
{\bf Figure 3}: Unfolded spectrum and best fit model (Table 1, Model 4),
shown in this figure as a solid line. The single components of the model are 
also shown, namely the MCD (dot-dashed line), the comptonized spectrum 
({\it Compst} model, dotted line), and the bremsstrahlung component
(dashed line). Also the absorption edge at 1.26 keV is visible.
\\ 

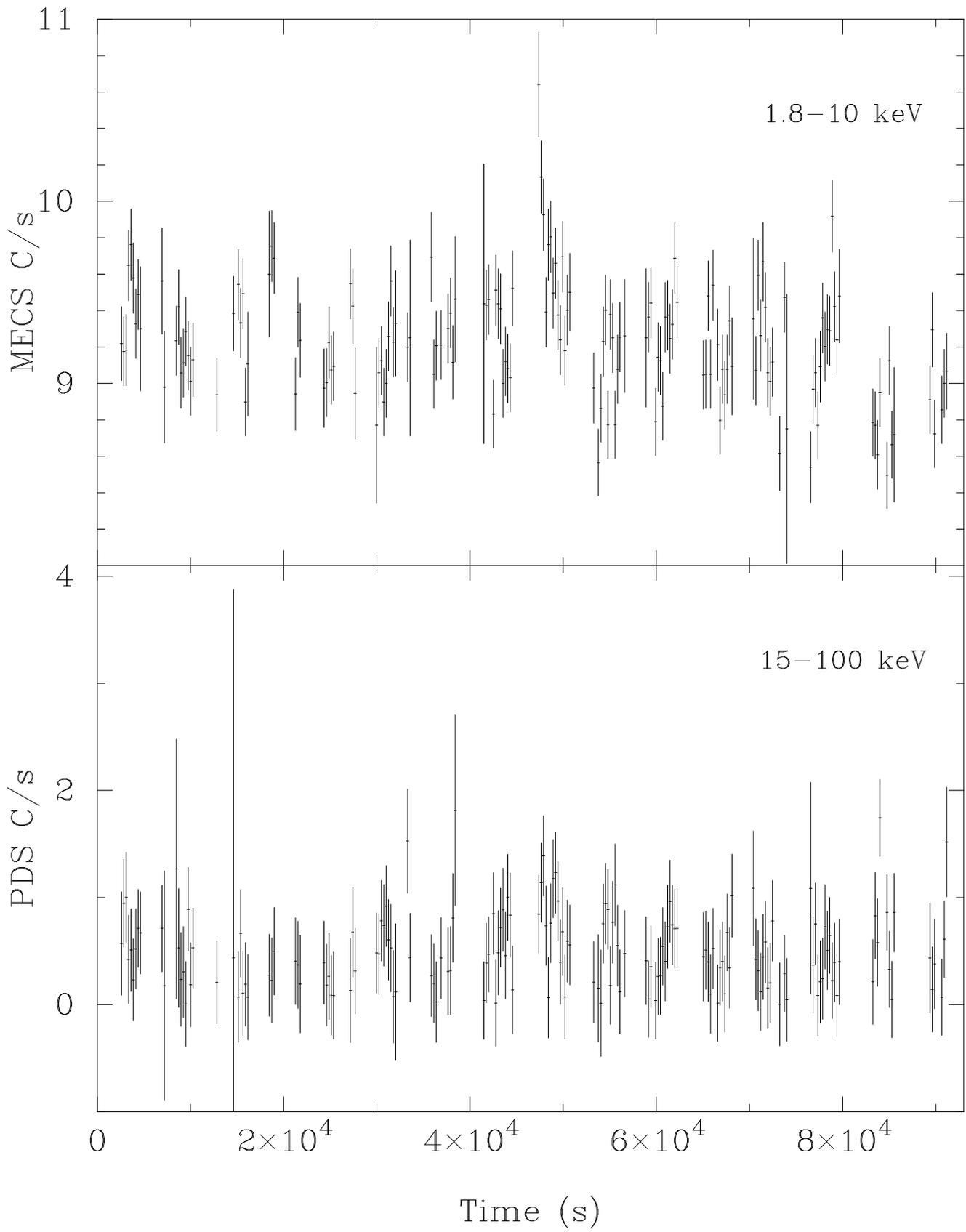
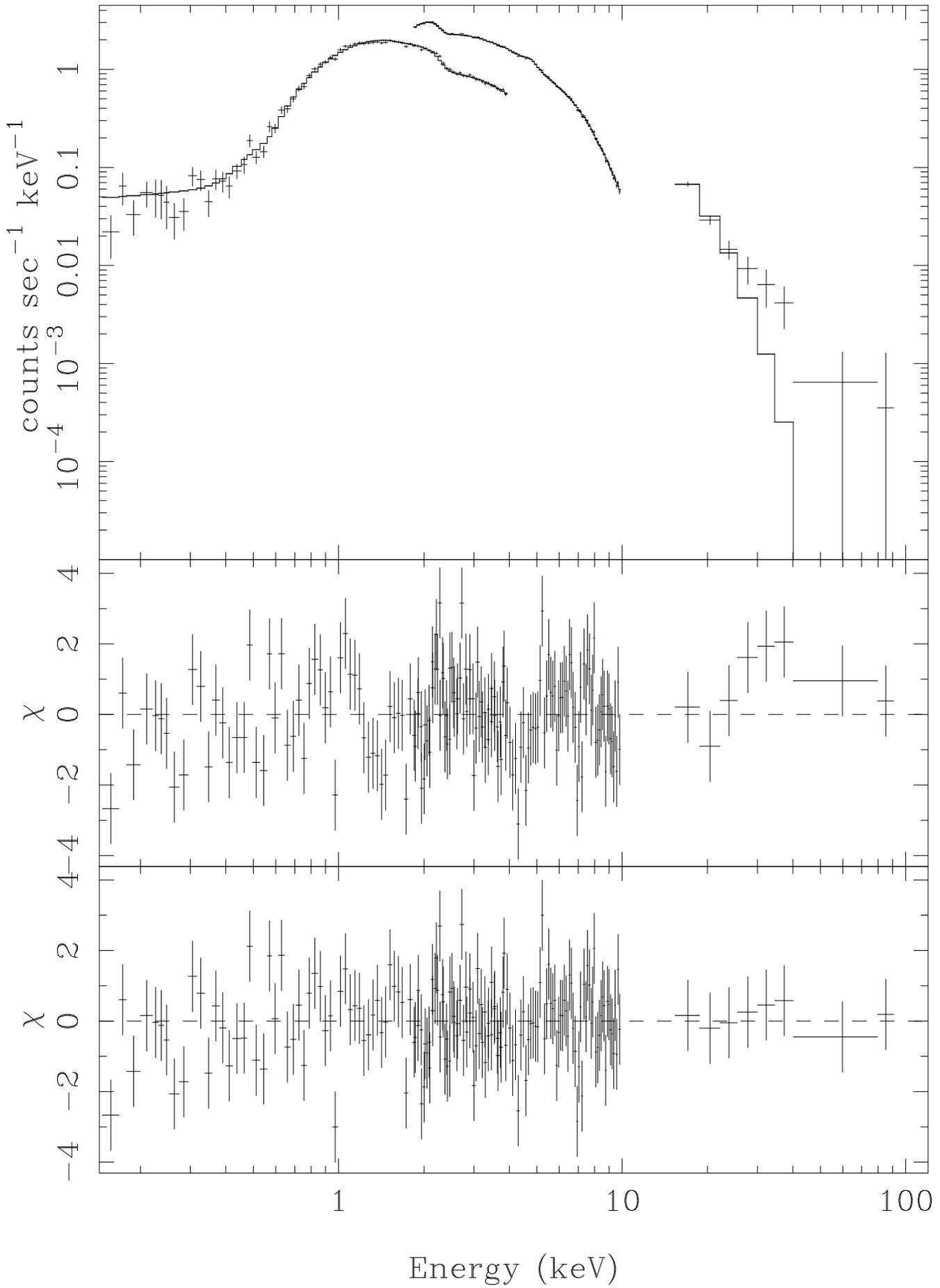
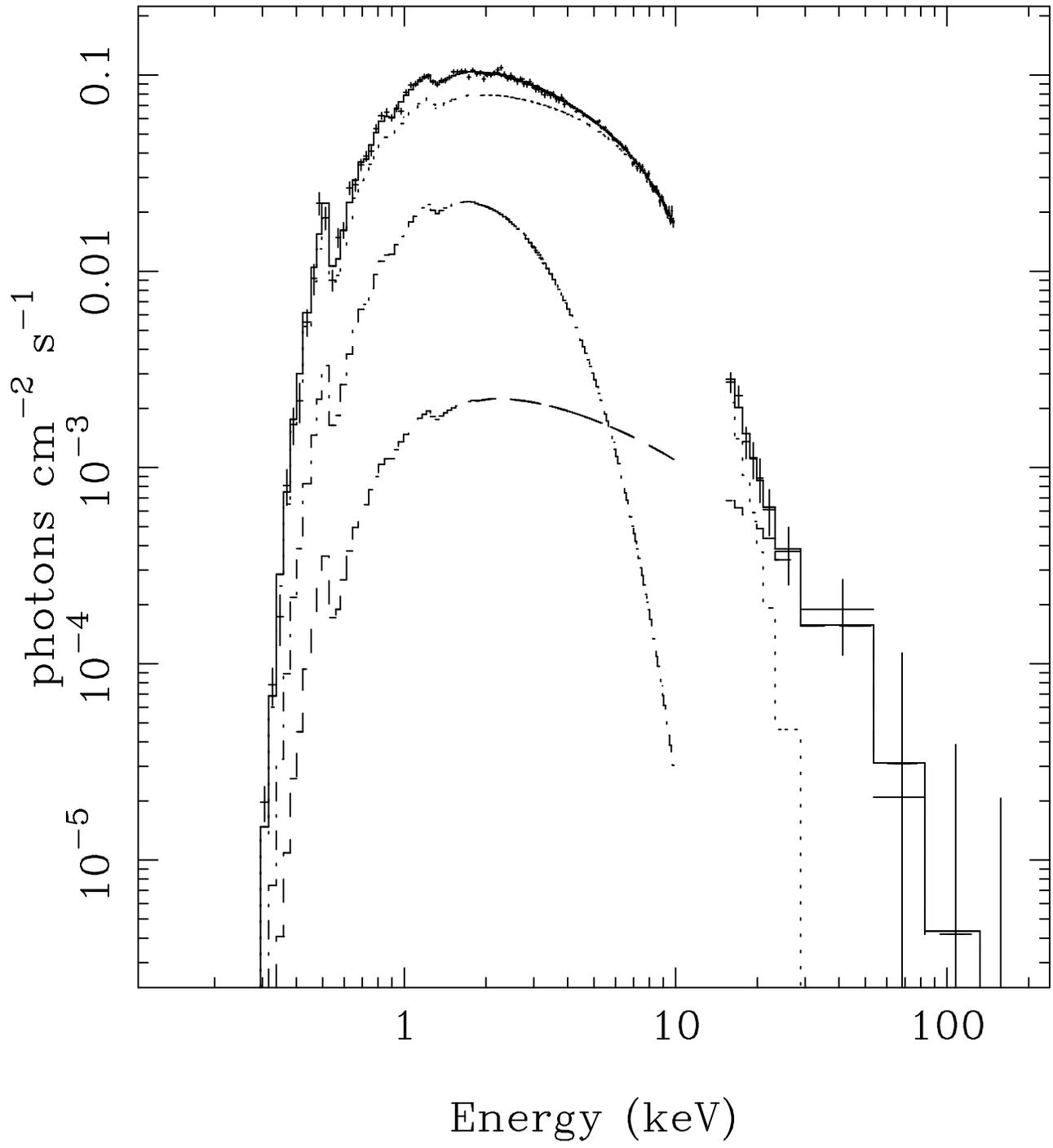
\end{document}